\documentstyle[aps,prl,preprint,epsf,floats]{revtex}
\tightenlines
\newcommand{\beq}{\begin{equation}}
\newcommand{\eeq}{\end{equation}}
\newcommand{\bqa}{\begin{eqnarray}}
\newcommand{\eqa}{\end{eqnarray}}

\def\square{\vcenter{\vbox{\hrule height.4pt
          \hbox{\vrule width.4pt height8pt
          \kern8pt\vrule width.4pt}\hrule height.4pt}}}

\def\sumint{\hbox{$\sum$}\!\!\!\!\!\!\int}
\def\isumint{\hbox{${\scriptstyle \Sigma}$}\!\!\!\!\int}

\begin{document}
\preprint{
\vbox{\halign{&##\hfil\cr
        & hep-ph/0002048 \cr
&\today\cr }}}

\title{The Massive Thermal Basketball Diagram}

\author{Jens O. Andersen and Eric Braaten}
\address{Physics Department, Ohio State University, Columbus OH 43210, USA}

\author{Michael Strickland}
\address{Physics Department, University of Washington, Seattle  WA 98195-1560}

\maketitle

\begin{abstract}
The ``basketball diagram'' is a three-loop vacuum diagram for 
a scalar field theory that cannot be expressed in terms of one-loop diagrams.
We calculate this diagram for a massive scalar field at nonzero temperature, 
reducing it to expressions involving three-dimensional integrals 
that can be easily evaluated numerically.
We use this result to calculate the free energy for a massive scalar field 
with a $\phi^4$ interaction to three-loop order. 
\end{abstract}

\newpage

\section{Introduction}

One of the obstacles to making progress in thermal field theory is that the technology
for explicit perturbative calculations is underdeveloped.  The formalism of
thermal field theory is sufficiently complicated that there are often
theoretical issues that are difficult to resolve without explicit
calculations.  An example is the gluon damping rate, which in the conventional
perturbative expansion is plagued by problems involving gauge invariance.  
A formal solution to the problem by the resummation of hard thermal loops 
was presented by Braaten and Pisarski in 1990 \cite{Braaten-Pisarski:HTL}.  
However, the solution was not widely accepted until the leading order expression 
for the damping rate was calculated explicitly \cite{Braaten-Pisarski:gammag}.

Around 1994, there was a significant step forward in the
calculational technology for massless theories.
The first perturbative calculation in thermal field theory 
that was carried out to high enough order that the running of the 
coupling constant came into play was a calculation of the free energy 
of a massless scalar field theory to order $g^4$ 
by Frenkel, Saa, and Taylor in 1992 \cite{F-S-T}.  
In 1994, there were several other calculations of the free energy
to fourth order in the coupling constant:
a calculation by Corian\`o and Parwani for QED \cite{Coriano-Parwani}
and completely analytic calculations by Arnold and Zhai \cite{Arnold-Zhai} 
for a massless scalar theory (correcting an error in Ref. \cite{F-S-T})
and for QCD.
Arnold and Zhai \cite{Arnold-Zhai} made a particularly significant 
contribution by showing 
how three-loop vacuum diagrams, such as the so-called ``basketball diagram" 
labeled 2b in Fig.~\ref{fig-F}, could be evaluated analytically.
These analytic calculations were then quickly extended to order $g^5$ 
for massless scalar theories \cite{Parwani-Singh,Braaten-Nieto:scalar},
abelian gauge theories \cite{Parwani,Andersen},
and nonabelian gauge theories \cite{Kastening-Zhai,Braaten-Nieto:QCD}.  
These explicit calculations revealed that the weak coupling expansion has 
convergence problems whose severity had not previously been appreciated.

Calculations in theories that include massive particles at 
nonzero temperature are more difficult, because there is a second scale 
in the problem.  The calculational technology for massive theories 
is much less well-developed than that for massless theories.
In addition to the obvious applications to theories with 
massive particles, calculations with massive propagators may 
also be useful for massless theories.  In such theories,
some of the most important thermal corrections have the 
effect of generating masses for the massless particles.
These corrections may be responsible for the poor 
convergence properties of the weak coupling expansion.
One of the most promising methods for resumming these corrections 
in scalar field theories is ``screened perturbation theory'' proposed
by Karsch, Patk\'os, and Petreczky \cite{K-P-P}.
This method involves adding and subtracting a mass term from the Lagrangian
and treating the subtracted term as a perturbation.
The integrals encountered in the screened perturbative expansion
are those of the corresponding massive theory.
Screened perturbation theory has been applied to the free energy 
at the two-loop level, and it seems to dramatically improve the 
convergence of the perturbative series  \cite{K-P-P}.
To determine how effective this method is in resumming the 
large perturbative corrections, it is essential to calculate 
higher order corrections explicitly.

In this paper, we take a step forward in the calculational technology 
for massive field theories by evaluating the basketball diagram.
The diagram cannot be evaluated analytically, but we reduce it to
expressions that involve integrals that are at most three-dimensional and can
easily be evaluated numerically.  Using our result for this diagram, we
calculate the free energy for the massive $\phi^4$ field theory through
next-to-next-to-leading order in the coupling constant $g^2$.

\section{Basketball Diagram}
\label{Basketball}

The {\it basketball diagram} is the diagram labeled 2b in Fig.~\ref{fig-F}.
At nonzero temperature $T$, it involves a three-fold
sum-integral over Euclidean momenta:
\begin{equation}
{\cal I}_{\rm ball} \;=\;
\sumint_{PQR}{1 \over P^2+m^2} {1 \over Q^2+m^2} {1 \over R^2+m^2}
	{1 \over (P+Q+R)^2+m^2}\;.
\label{basketball}
\end{equation}
The Euclidean four-momentum is $P = ({\bf p},p_4 = 2 \pi n T)$, 
where $n$ is an integer, and its square is $P^2 = {\bf p}^2 + p_4^2$.
The sum-integral $\isumint_P$ represents the sum over the Euclidean energies 
and the integral over the spatial momentum:
\begin{equation}
\sumint_P \;=\;
T \sum_{p_4} \, \int_{\bf p} \,.
\end{equation}

We use dimensional regularization of the integral over ${\bf p}$
to cut off the ultraviolet divergences 
in the sum-integral.  Our convention for the measure in the integral is
\begin{equation}
\int_{\bf p} \;=\;
\left( {e^\gamma \mu^2 \over 4\pi} \right)^\epsilon 
	\, \int {d^{3-2\epsilon}p \over (2 \pi)^{3-2\epsilon}} \,,
\end{equation}
where $3-2 \epsilon$ is the number of spatial dimensions,
$\gamma$ is Euler's constant,  and $\mu$ is an arbitrary
momentum scale.  The factor of $\mu^{2 \epsilon}$ gives
the dimensionally-regularized integral (\ref{basketball}) dimensions 
of (energy)$^4$.

\subsection{Decomposition into integrals}

In order to evaluate the sum-integral (\ref{basketball}), we first reduce it to
integrals that contain factors of the Bose-Einstein distribution function 
$n(E)= 1/(e^{\beta E} -1)$ with positive energy $E$.  
We use the method of Bugrij and Shadura \cite{Bugrij-Shadura} 
which expresses the coefficients of the
Bose-Einstein factors in terms of S-matrix elements at zero temperature.  
This strategy was also used by Frenkel, Taylor and Saa \cite {F-S-T} 
to evaluate the massless basketball diagram.  Although the derivation 
by Bugrij and Shadura is lengthy, their final result can be obtained 
by making some simple substitutions in the expression (\ref{basketball}).
The sum-integrals over Euclidean momenta $P = ({\bf p}, p_4)$
are replaced by integrals over Minkowski momenta $p = (p_0, {\bf p})$:
$\isumint_P \longrightarrow -i \int_p$.
The Euclidean propagators are replaced by the Minkowski propagators
of the real-time formalism:
\begin{equation}
{1 \over P^2+m^2} \;\longrightarrow\;
i \left( {i \over p^2-m^2 + i \varepsilon} 
	+ n(|p_0|) 2 \pi \delta(p^2-m^2) \right),
\end{equation}
where $p^2 = p_0^2 - {\bf p}^2$.  
The sum-integral (\ref{basketball}) is then
given by the real part of the resulting 
expression, which now involves a three-fold integral over Minkowski momenta.
For some of the integration momenta, the energy appears in the argument of a
Bose-Einstein distribution.  It is convenient to Wick-rotate the remaining
integration momenta back to Euclidean space:
$\int_p \longrightarrow i \int_P$,
where the measure of the dimensionally regularized integral over  
the Euclidean momentum $P$ is
\begin{equation}
\int_P \;=\;
\left( {e^\gamma \mu^2 \over 4\pi} \right)^\epsilon 
	\, \int {d^{4-2\epsilon}p \over (2 \pi)^{4-2\epsilon}} \,.
\end{equation}
Having carried out this procedure, 
the basketball diagram can be expressed as
the sum of terms with different numbers of Bose-Einstein factors:
\begin{equation}
I_{\rm ball} \;=\; 
{\cal I}^{(0)} + 4 \, {\cal I}^{(1)} + 6 \, {\cal I}^{(2)} 
	+ 4 \, {\cal I}^{(3)} \, .
\label{I-0123}
\end{equation}
The term with four Bose-Einstein factors is purely imaginary 
and it vanishes when we take the real part.
The other terms are
\begin{eqnarray}
{\cal I}^{(0)} &=&
\int_{PQR}{1 \over P^2+m^2} {1 \over Q^2+m^2} {1 \over R^2+m^2}
	{1 \over (P+Q+R)^2+m^2} \;,
\label{I0-def}
\\
{\cal I}^{(1)} &=&
\int_p n \delta(p)
\int_{QR}{1 \over Q^2+m^2} {1 \over R^2+m^2}
	{1 \over (P+Q+R)^2+m^2}\Bigg|_{P^2 = -m^2} \;,
\label{I1-def}
\\
{\cal I}^{(2)} &=&
{\rm Re}\int_p n \delta(p) \int_q n \delta(q)
\int_R {1 \over R^2+m^2}
{1 \over (P+Q+R)^2+m^2} \Bigg|_{(P+Q)^2 = -(p+q)^2 - i \varepsilon} \;,
\label{I2-def}
\\
{\cal I}^{(3)} &=&
{\rm Re}\int_p n \delta(p) \int_q n \delta(q)  \int_r n\delta(r)
	 {(-1) \over (p+q+r)^2-m^2 + i \varepsilon} \;,
\label{I3-def}
\end{eqnarray}
where we have used the shorthand 
$n\delta(p) = n(|p_0|) 2 \pi \delta (p^2 - m^2)$.

\subsection{Zero thermal factors}

We first consider the term ${\cal I}^{(0)}$, 
which is the integral for the zero-temperature basketball
diagram.  The poles in $\epsilon$ for this diagram have been calculated by
Kastening \cite{Kastening} and by Chung and Chung \cite{Chung-Chung}.  
The finite terms in the diagram can be obtained by following the 
strategy used in Appendix B of Ref. \cite{Braaten-Nieto:scalar} 
to calculate the zero-temperature basketball diagram in three dimensions.

The diagram is first Fourier-transformed to coordinate space,
which reduces it to an integral over a single coordinate $R$:
\begin{equation}
{\cal I}^{(0)} = \int_R V^4 (R) \, ,
\end{equation}
where the potential $V(R)$ is
\begin{equation}
V(R) \;=\; \left( {e^\gamma \mu^2 \over 4\pi} \right)^\epsilon
{1 \over (2 \pi)^{2-\epsilon}} \left( {m \over R} \right)^{1-\epsilon}
K_{1-\epsilon}(mR) 
\end{equation}
and $K_\nu(z)$ is a modified Bessel function.  
The measure for the integration over $R$ is 
\begin{equation}
\int_R \;=\; \left( {e^\gamma \mu^2 \over 4\pi} \right)^{-\epsilon} 
\int d^{4-2\epsilon}R \, .
\end{equation}
After integrating over angles in $4-2\epsilon$ dimensions,
the integral reduces to
\begin{equation}
{\cal I}^{(0)} \;=\; 
{m^4 \over (4 \pi)^6} \left( {e^\gamma \mu^2 \over m^2} \right)^{3\epsilon}
{32 \over \Gamma(2-\epsilon)} 
\int_0^\infty dt ~ t^{-1+2 \epsilon} K_{1-\epsilon}^4(2t) \, .
\label{I0-int} 
\end{equation}
The $t \to 0$ region of the integral gives poles in $\epsilon$.
The small-$t$ behavior of the Bessel function is given by the 
power-series expansion
\begin{eqnarray}
K_{1-\epsilon}(2t) =
{\Gamma (1 - \epsilon) t^{-1 + \epsilon} \over 2}
+ {\Gamma (1 - \epsilon) \Gamma(\epsilon) \over 2} \sum^\infty_{j = 0} 
\left( {t^{2j+1+\epsilon} \over (j+1)! \Gamma(j+1+\epsilon)}
	- {t^{2j+1-\epsilon} \over j! \Gamma(j+2-\epsilon)} \right)  .
\label{K-ps}
\end{eqnarray}
In the integral over $t$ in (\ref{I0-int}), 
the poles in $\epsilon$ come from the $t^{-5+6\epsilon}$, $t^{-3+6\epsilon}$, 
$t^{-3 + 4 \epsilon}$, $t^{-1 + 6 \epsilon}$, 
$t^{-1 + 4 \epsilon}$, and $t^{-1 + 2 \epsilon}$ terms.  
We can calculate the poles analytically by
multiplying each of these terms by an appropriate convergence factor and
integrating over $t$.  After these terms, with their convergence factors, are
subtracted from the original integrand, the remaining integral is convergent 
for $\epsilon = 0$ and can be evaluated numerically.  
We choose convergence factors of the form $(e^{8t})_n e^{-8t}$,
where $(e^x)_n$ is the truncated power series for the exponential function:
$(e^x)_n = \sum^n_{i = 0}x^i/i!$.  This convergence factor
behaves like $1 + {\cal O} (t^{n+1})$ at small $t$, and 
has the same exponential falloff as $K_1^4(2t)$ at large $t$.
The resulting expression for the integral is in the limit $\epsilon \rightarrow 0$
\begin{eqnarray}
&& \int^\infty_0 dt ~ t^{-1 + 2 \epsilon} K^4_{1-\epsilon}(2t) 
\nonumber
\\
&& \;=\; {\Gamma^4(1-\epsilon) \over 16} 
\int^\infty_0 dt ~ t^{-1+2\epsilon} e^{-8t}
\Bigg[ t^{-4+4\epsilon} \left( e^{8t} \right)_4
+ {4  t^{-2+2\epsilon} \over \epsilon} \left( t^{2\epsilon} 
	- {\Gamma(1+\epsilon) \over \Gamma(2-\epsilon)} \right)
	\left( e^{8t} \right)_2 
\nonumber
\\
&& \qquad \qquad \qquad \qquad \qquad
+ {2t^{2 \epsilon} \over \epsilon (1+\epsilon)} 
\left( t^{2 \epsilon} 
	- {2 \Gamma(2+\epsilon) \over \Gamma(3-\epsilon)} \right)
+ {6 \over \epsilon^2}
\left( t^{2 \epsilon} - {\Gamma(1+\epsilon) \over \Gamma(2-\epsilon)} \right)^2
	\Bigg]
\nonumber
\\
&&  \qquad \qquad
\;+\; \int^\infty_0 dt ~ {1 \over t} 
\Bigg\{ K_1^4(2t) 
- {e^{-8t} \over 16} \bigg [{1 \over t^4} (e^{8t})_4 
	+ {4 \over t^2} (2 \log t + 2 \gamma - 1) (e^{8t})_2
\nonumber
\\ 
&& \qquad \qquad \qquad  \qquad \qquad \qquad \qquad
+ 2 (2 \log t + 2 \gamma - \mbox{${5 \over 2}$} ) 
+ 6 ( 2 \log t + 2 \gamma - 1 )^2  \bigg] \Bigg\} \, .
\label{int-K4}
\end{eqnarray}
The first integral in (\ref{int-K4}) can be evaluated 
analytically in terms of gamma functions, 
and it reduces in the limit $\epsilon \to 0$ to
\begin{eqnarray} 
\Gamma(1+6 \epsilon) && 
\Bigg\{ {1 \over \epsilon^3} + {17 \over 6 \epsilon^2}
+ {59 - 36 \pi^2 \over 12\epsilon} - {821 \over 24} + 381 \log 2 
\nonumber
\\
&& - 90 \log^2 2 - 216 \log^3 2
- {61 \pi^2 \over 6} - 12 \pi^2 \log 2 - 27 \psi''(1) \Bigg\} \,,
\end{eqnarray}
where $\psi(z)$ is the digamma function.
The numerical value of the second integral in (\ref{int-K4}) is 
0.36106. 
Inserting these results into (\ref{I0-int}) and keeping terms through order
$\epsilon^0$, our final result is
\begin{equation} 
{\cal I}^{(0)} \;=\;
{1 \over (4\pi)^6} \left( {\mu\over m} \right)^{6\epsilon}
\left[ {2 \over \epsilon^3} + {23 \over 3 \epsilon^2}
	+ {35+\pi^2 \over 2 \epsilon} + C_0 \right] m^4 \,,
\label{I0-final}
\end{equation}
where the numerical value of the constant is $C_0 = 39.429$.  


\subsection{One thermal factor}

The expression (\ref{I1-def}) for ${\cal I}^{(1)}$ can be written as
\begin{equation} 
{\cal I}^{(1)} \;=\; 
I_{\rm sun}(-m^2) \int_p n(|p_0|) 2 \pi \delta(p^2-m^2) \,,
\label{I1-prod}
\end{equation}
where $I_{\rm sun}(P^2)$ is the integral for the ``setting sun diagram'' 
in the boson self-energy at zero temperature:
\begin{equation} 
I_{\rm sun}(P^2) \;=\;
\int_{QR}{1 \over Q^2+m^2} {1 \over R^2+m^2} {1 \over (P+Q+R)^2+m^2} \,.
\label{I-sun}
\end{equation}
The integral over $p$ in (\ref{I1-prod}) can be written as
\begin{equation} 
\int_p n(|p_0|) 2 \pi \delta(p^2-m^2) \;=\;
{1 \over (4 \pi)^2}  \left( {\mu \over m} \right)^{2\epsilon} J_1 T^2\;,
\label{int-deltan}
\end{equation}
where $J_1$ is a function of $\beta m$ defined by (\ref{Jn}) in the Appendix. 

The setting-sun integral (\ref{I-sun}) at $P^2 = - m^2$ 
can be evaluated by following the 
strategy used in Appendix B of Ref. \cite{Braaten-Nieto:scalar} 
to calculate the corresponding integral in three dimensions.
After Fourier-transforming, 
it reduces to an integral over a single coordinate $R$:
\begin{equation} 
I_{\rm sun}(-m^2) \;=\; 
\int_R e^{i P \cdot R} V^3(R)\Bigg|_{P^2 = -m^2}  \;.
\label{Isun-R}
\end{equation}
After averaging over angles in 
$4-2\epsilon$ dimensions and evaluating at $P^2 = - m^2$,
the exponential factor becomes
\begin{equation} 
\left\langle e^{i P \cdot R} \right\rangle\Bigg|_{P^2 = -m^2} \;=\; 
\Gamma(2-\epsilon) \left({2 \over m R} \right)^{1-\epsilon} 
	I_{1-\epsilon}(mR) \,, 
\end{equation}
where $I_\nu(z)$ is a modified Bessel function.
The integral over $R$ then reduces to a one-dimensional integral,
and (\ref{Isun-R}) becomes
\begin{equation}
I_{\rm sun}(-m^2) \;=\; 
{m^2 \over (4 \pi)^4} \left( {\mu \over m} \right)^{4\epsilon} 
16 e^{2 \gamma \epsilon}
\int_0^\infty dt ~ t^{-1+2 \epsilon} I_{1-\epsilon}(2t) K_{1-\epsilon}^3(2t).
\label{Isun-int} 
\end{equation}
The $t \to 0$ region of the integral gives poles in $\epsilon$.  
The small-$t$ behavior of the Bessel function $I_{1 - \epsilon}(2t)$ 
is given by the power-series expansion
\begin{eqnarray}
I_{1 - \epsilon}(2t) \;=\;
\sum^\infty_{j = 0} {t^{2 j + 1 - \epsilon} \over j! \Gamma(j+2-\epsilon)}\,.
\end{eqnarray}
In the integral over $t$ in (\ref{Isun-int}), 
the poles in $\epsilon$ come from the $t^{-3 + 4 \epsilon}$, 
$t^{-1 + 4 \epsilon}$, and $t^{-1 + 2 \epsilon}$ terms.
We can calculate the poles analytically by
multiplying each of these terms by an appropriate convergence factor and
integrating over $t$.  After these terms, with their convergence factors, are
subtracted from the original integrand, the remaining integral is convergent 
for $\epsilon = 0$ and can be evaluated numerically.  
We choose convergence factors of the form $(e^{6t})_n e^{-6t}$.
The resulting 
expression for the integral is in the limit $\epsilon \rightarrow 0$

\begin{eqnarray}
&&\int^\infty_0 dt ~ t^{-1 + 2 \epsilon} ~ 
	I_{1-\epsilon}(2t) ~ K^3_{1-\epsilon}(2t) 
\nonumber 
\\
&& \;=\; {\Gamma^2(1-\epsilon) \over 8(1-\epsilon)} 
\int^\infty_0 dt ~ t^{-1+2\epsilon}e^{-6t}
\Bigg[ t^{-2+2\epsilon} \left( e^{6t} \right)_2
+ {t^{2\epsilon} \over 2-\epsilon}  
+ {3 \over \epsilon } 
\left( t^{2 \epsilon} 
	- {\Gamma(1+\epsilon) \over \Gamma(2-\epsilon)} \right)
	\Bigg]
\nonumber
\\
&&  \qquad \qquad
\;+\; \int^\infty_0 dt ~ {1 \over t} 
\Bigg\{ I_1(2t) K_1^3(2t) 
- {e^{-6t} \over 8} \bigg[ {1 \over t^2} (e^{6t})_2 
	+ {1 \over 2} + 3(2 \log t + 2 \gamma - 1)  \bigg] \Bigg\} \,.
\label{int-IK3}
\end{eqnarray}
The first integral in (\ref{int-IK3}) can be evaluated 
analytically in terms of gamma functions, 
and it reduces in the limit $\epsilon \to 0$ to
\begin{eqnarray} 
\Gamma(1+4 \epsilon)
\Bigg\{ - {3 \over 4 \epsilon^2}
- {11 \over 8 \epsilon} - {167 \over 16} + {5 \over 2} \log 6 
+ 3 \log^2 6 + {3 \pi^2 \over 2} \Bigg\} \,.
\end{eqnarray}
The numerical value of the second integral in (\ref{int-IK3}) is 
$-1.2713$. 
Inserting these results into (\ref{Isun-int}) and keeping terms through order
$\epsilon^0$, we obtain
\begin{eqnarray} 
I_{\rm sun}(-m^2) &=&
{1 \over (4\pi)^4} \left( {\mu\over m} \right)^{4\epsilon}
\left[ -{3 \over 2\epsilon^2} - {17 \over 4 \epsilon} + C_1 \right] 
	m^2 \;,
\label{Isun-final}
\end{eqnarray}
where the numerical value of the constant is $C_1 = -9.8424$.
Inserting (\ref{Isun-final}) and (\ref{int-deltan}) into (\ref{I1-prod}),
our final result is
\begin{eqnarray} 
{\cal I}^{(1)} &=&
{1 \over (4\pi)^6} \left( {\mu\over m} \right)^{6\epsilon}
\left[ -{3 \over 2\epsilon^2} - {17 \over 4 \epsilon} + C_1 \right] 
	J_1 m^2 T^2\;.
\label{I1-final}
\end{eqnarray}
Note that the integral $J_1$ depends on $\epsilon$.

\subsection{Two thermal factors}

The expression (\ref{I2-def}) for ${\cal I}^{(2)}$ involves the 
``bubble integral'' 
\begin{eqnarray} 
I_{\rm bubble}(P^2) \;=\; \int_R {1 \over R^2+m^2} {1 \over (P+R)^2+m^2} \,,
\label{Ibubble}
\end{eqnarray}
which can be evaluated using a Feynman parameter:
\begin{eqnarray} 
I_{\rm bubble}(P^2) \;=\; 
{1 \over (4\pi)^2} \left( {\mu\over m} \right)^{2\epsilon}
\left[ {1 \over \epsilon} 
	- \int_0^1 dx \log {m^2 + x(1-x)P^2 \over m^2}  
	\right] \;.
\label{feynman}
\end{eqnarray}
The real part of the integral evaluated at 
$P^2 = - (p+q)^2 - i \varepsilon$
is obtained by simply replacing the argument of the logarithm 
by its absolute value.
When (\ref{feynman}) is inserted into (\ref{I2-def}), 
the coefficient of $1/\epsilon$ can be evaluated using (\ref{int-deltan}).
Reducing the other term to an integral over spatial momenta,
we obtain
\begin{eqnarray} 
{\cal I}^{(2)} &=& {1 \over (4\pi)^6} \left( {\mu\over m} \right)^{6\epsilon}
\Bigg\{ {1 \over \epsilon} J_1^2 T^4 
- 32 
\int_0^\infty dp \, {p^2  n(E_p) \over E_p}
	\int_0^\infty dq \, {q^2 n(E_q) \over E_q} 
\nonumber
\\ 
&& \hspace{5cm}
\times \sum_\sigma \int_0^1 dx \left\langle
\log {|m^2 - x(1-x)(E_\sigma^2 - k^2)| \over m^2} 
	\right\rangle \Bigg\} \,,
\end{eqnarray}
where $E_\sigma = E_p + \sigma E_q$, $k = |{\bf p} + {\bf q}|$,
the sum is over $\sigma = \pm$, and
the angular brackets denote the
average over the angles of ${\bf p}$ and ${\bf q}$.
It is convenient to change the angular integration variable to $k$.  
The integral over $x$ in the angular average then reduces to
\begin{eqnarray} 
\int_0^1 dx \left\langle
\log {|m^2 - x(1-x)(E_\sigma^2 - k^2)| \over m^2} \right\rangle
\;=\; {1 \over 2 p q} \int_{|p-q|}^{p+q} dk ~ k ~ 
	\left[ f_2(E_\sigma,k) - 2 \right] \,,
\end{eqnarray}
where the function in the integrand is
\begin{eqnarray} 
f_2(E,k) &=& 
\left( {E^2-M_k^2 \over E^2-k^2} \right)^{1/2}
\log { (E^2-k^2)^{1/2} + (E^2-M_k^2)^{1/2} \over
	(E^2-k^2)^{1/2} - (E^2-M_k^2)^{1/2} } \, ,
\hspace{1cm}  k^2 < E^2 - 4 m^2 \, ,
\nonumber
\\
&=& 
2 \left( {M_k^2-E^2 \over E^2-k^2} \right)^{1/2} 
	{\rm atan} \left( {E^2-k^2 \over M_k^2-E^2} \right)^{1/2} \, ,
\hspace{3cm} E^2-4 m^2 < k^2 < E^2 \, ,
\nonumber
\\
&=& 
\left( {M_k^2-E^2 \over k^2-E^2} \right)^{1/2}
\log { (M_k^2-E^2)^{1/2} + (k^2-E^2)^{1/2} \over
	(M_k^2-E^2)^{1/2} - (k^2-E^2)^{1/2} } \, ,
\hspace{1cm} E^2 < k^2 \, .
\end{eqnarray}
and $M_k^2 = 4 m^2 + k^2$. Our final result is 
\begin{eqnarray} 
{\cal I}^{(2)} &=& {1 \over (4\pi)^6} \left( {\mu\over m} \right)^{6\epsilon}
\left[ \left( {1 \over \epsilon} + 2 \right) J_1^2 + K_2 \right] T^4\;,
\label{I2-final}
\end{eqnarray}
where $K_2$ is the function of $\beta m$ defined by the following integral: 
\begin{eqnarray} 
K_2 &=& - {32 \over T^4} \int_0^{\infty} dp \; {p n(E_p) \over E_p} 
\int_0^p dq \; {q n(E_q) \over E_q}  \int_{p-q}^{p+q} dk \; k \;
	\sum_\sigma f_2(E_\sigma,k) \;.
\label{K2}
\end{eqnarray}

In the limit $m \to 0$, $K_2$ reduces to
\begin{equation} 
K_2 \;\longrightarrow\; 
- {(4 \pi)^4 \over 72}
\left[ \log{4 \pi T \over m} - {1 \over 2} 
	-  {\zeta'(-1) \over \zeta(-1)} \right]  \,,
\label{K2-zero}
\end{equation}
where $\zeta(z)$ is the Riemann zeta function.
In that same limit, $J_1$ reduces to 
\begin{equation} 
J_1 \;\longrightarrow\; 
{(4 \pi)^2 \over 12}
\left\{ 1 + 2 \epsilon \left[ - \log{4 \pi T \over m} + 1 
	+  {\zeta'(-1) \over \zeta(-1)} \right] 
+ O(\epsilon^2) \right\} \,.
\label{J1-zero}
\end{equation}
Inserting (\ref{K2-zero}) and (\ref{J1-zero}) into (\ref{I2-final}),
we obtain the expression for ${\cal I}^{(2)}$ in the limit $m \to 0$:
\begin{equation} 
{\cal I}^{(2)} \;\longrightarrow\; 
{1 \over 144 (4 \pi)^2} \left( {\mu \over 4 \pi T} \right)^{6\epsilon}
\left[ {1 \over \epsilon} + 7 + 6 {\zeta'(-1) \over \zeta(-1)} \right] \,.
\label{I2-zero}
\end{equation}
This agrees with the analytic result first obtained
by Frenkel, Saa, and Taylor \cite{F-S-T}.

\subsection{Three thermal factors}

The integral ${\cal I}^{(3)}$ in (\ref{I3-def}) is finite 
in three spatial dimensions, so we can set $\epsilon = 0$ from the beginning.
After using the delta functions to integrate over $p_0$, $q_0$, and $r_0$,
the integral reduces to
\begin{eqnarray} 
{\cal I}^{(3)} &=& {128 \over (4\pi)^6} 
\int_0^\infty dp \, {p^2 n(E_p) \over E_p} 
\int_0^\infty dq \, {q^2 n(E_q) \over E_q}
\int_0^\infty dr \, {r^2 n(E_r) \over E_r} 
\sum_{\sigma,\tau} \left\langle  {\cal P}
{(-1) \over E_{\sigma\tau}^2 - k^2 - m^2} \right\rangle \,,
\end{eqnarray}
where $E_{\sigma\tau} = E_p + \sigma E_q + \tau E_r$, 
$k =|{\bf p} + {\bf q} + {\bf r}|$,
the sum is over $\sigma=\pm$ and $\tau=\pm$,
the angular brackets denote the
average over the angles of ${\bf p}$, ${\bf q}$,  and ${\bf r}$,
and $ {\cal P}$ denotes the principal value prescription
for the poles in the propagator.
Before averaging over the angles of ${\bf p}$, ${\bf q}$, and ${\bf r}$,
it is convenient to use the symmetry in the integration variables 
to impose the restriction $r < q < p$ while multiplying by $3!$.
We can then average over angles using the identity
\begin{equation} 
\left\langle  F(|{\bf p} + {\bf q} + {\bf r}|) 
	\right\rangle \;=\; 
{1 \over 4 p q r} \int_0^{p+q+r} dk ~ k ~ w(p,q,r,k) ~ F(k) \,,
\end{equation}
where the weight function for the case $r<q<p$ is 
\begin{eqnarray} 
w(p,q,r,k) 
&=& 2k ~ \theta(q+r-p) \, ,      \hspace{1.5cm} 0 < k < |p-q-r| \, ,
\nonumber
\\
&=& k+q+r-p \, , \hspace{2cm} |p-q-r| < k < p+r-q \, ,
\nonumber
\\
&=& 2r \, ,      \hspace{4cm} p+r-q < k < p+q-r \, ,
\nonumber
\\
&=& p+q+r-k \, , \hspace{2cm} p+q-r < k < p+q+r \, .
\end{eqnarray}
Integrating over $k$, our final result is
\begin{equation} 
{\cal I}^{(3)} \;=\; {1 \over (4\pi)^6} K_3 T^4\,,
\label{I3-final}
\end{equation}
where $K_3$ is the function of $\beta m$ 
defined by the following integral:
\begin{eqnarray} 
K_3 \;=\; 
{96 \over T^4} \int_0^{\infty} dp \; {p n(E_p) \over E_p} 
\int_0^p dq \; {q n(E_q) \over E_q} \int_{0}^{q} dr \; 
	{r n(E_r) \over E_r} 
\sum_{\sigma\tau} \left[ f_3(E_{\sigma\tau},p+q+r) \right. 
\nonumber
\\ 
\left. 
- f_3(E_{\sigma\tau},p+q-r) - f_3(E_{\sigma\tau},p-q+r)
+ f_3(E_{\sigma\tau},p-q-r) \right] \;.
\label{K3}
\end{eqnarray}
The function in the integrand is
\begin{eqnarray} 
f_3(E,p) &=& p \log{m^2-E^2+p^2 \over m^2}
+ 2 (m^2-E^2)^{1/2} \; {\rm atan}{p \over (m^2-E^2)^{1/2}}\;,
\hspace{1.13cm} E^2 < m^2 \;,
\nonumber
\\
&=& p \log{|E^2-m^2-p^2| \over m^2} 
+ (E^2-m^2)^{1/2} \; \log{ (E^2-m^2)^{1/2}+p \over |(E^2-m^2)^{1/2}-p| }\;,
\hspace{0.3cm}E^2>m^2\;.
\end{eqnarray}

The integral $K_3$ in the limit $m \to 0$ was calculated 
by Frenkel, Saa, and Taylor numerically \cite{F-S-T}
and by Arnold and Zhai analytically \cite{Arnold-Zhai}:
\begin{equation} 
K_3 \;\longrightarrow\; 
{(4 \pi)^4 \over 48}
\left[ - {7 \over 15} +  {\zeta'(-1) \over \zeta(-1)}
		- {\zeta'(-3) \over \zeta(-3)} \right] \,.
\end{equation}
Its numerical value in this limit is $K_3 \to 453.51$.  

\subsection{Total}

The final result for the basketball diagram is obtained by inserting
(\ref{I0-final}), (\ref{I1-final}), (\ref{I2-final}), and (\ref{I3-final}) 
into (\ref{I-0123}):
\begin{eqnarray} 
{\cal I}_{\rm ball}
&=& {1 \over (4\pi)^6} \left( {\mu\over m} \right)^{6\epsilon}
\Bigg\{ \left[ {2 \over \epsilon^3} + {23 \over 3 \epsilon^2}
	+ {35+\pi^2 \over 2 \epsilon} + C_0 \right] m^4  
+ \left[ - {6 \over \epsilon^2} - {17 \over \epsilon} + 4 C_1 \right] 
	J_1 m^2 T^2
\nonumber 
\\
&& \hspace{3cm}
+ \left( {6 \over \epsilon} + 12 \right) J_1^2 T^4
	+ (6 K_2 + 4 K_3) T^4 \Bigg\}\;.
\label{Iball-final}
\end{eqnarray}
To obtain the Laurent expansion including all terms through order $\epsilon^0$,
it remains only to expand the factor $(\mu/m)^{6\epsilon}$ and the integral $J_1$
in powers of $\epsilon$.
In the limit $m \to 0$, (\ref{Iball-final}) reduces to the analytic result obtained by
Arnold and Zhai \cite{Arnold-Zhai}:
\begin{eqnarray} 
{\cal I}_{\rm ball} &\longrightarrow& 
{1 \over 24 (4\pi)^2} \left( {\mu\over 4 \pi T} \right)^{6\epsilon}
\left[ {1 \over \epsilon} + {91 \over 15} +  8 {\zeta'(-1) \over \zeta(-1)}
		- 2 {\zeta'(-3) \over \zeta(-3)} \right] \,.
\label{Iball-zero}
\end{eqnarray}

\section{Three-Loop Free Energy}
\label{Free Energy}

Using our result for the massive basketball diagram, 
we can calculate the free energy for a 
massive scalar field theory with a $\phi^4$ interaction to three-loop order.
The Lagrangian for the field theory is
\bqa
{\cal L} \;=\; {1\over2} \partial_{\mu}\phi\partial^{\mu}\phi 
- {1 \over 2} {\bar m}^2 \phi^2
- {1\over 24} {\bar g}^2 \mu^{2 \epsilon} \phi^4 
+ \Delta{\cal L}\;,
\label{barel}
\eqa
where $\Delta{\cal L}$ includes the counterterms. 
We define the parameters ${\bar m}$ and ${\bar g}$ by dimensional
regularization and minimal subtraction, so they depend implicitly on  
the renormalization scale $\mu$.

\begin{figure}
\epsfxsize=9cm
\centerline{\epsffile{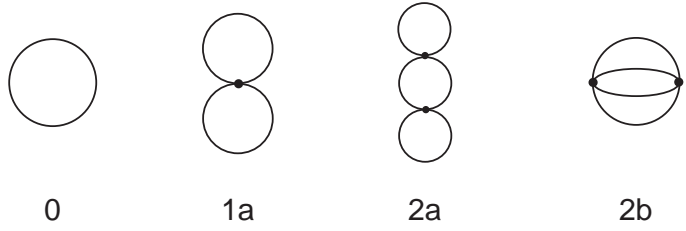 }}
\caption{One-loop, two-loop, and three-loop diagrams 
	contributing to the free energy.} 
\label{fig-F}
\end{figure}

\subsection{One loop}\label{1.1}

The free energy at zeroth order in $\bar g$ is given by the one-loop
diagram labeled 0 in Fig.~\ref{fig-F}:
\bqa
{\cal F}_0 &=& -{1\over2}{\cal I}_0' \,,
\label{F0-int}
\eqa
where the sum-integral ${\cal I}_0'$ is defined by (\ref{I0'})
in the Appendix.
Keeping only the temperature-dependent term, 
the result for the one-loop contribution to the free energy is 
\bqa
{\cal F}_0 &=&
- {1 \over 2(4\pi)^2} \left( {\mu\over\bar m} \right)^{2\epsilon}
	J_0 T^4 \;,
\label{F0}
\eqa
where $J_0$ is the function of $\beta \bar m$ defined in (\ref{Jn}).
In the limit $\epsilon \to 0$, this function reduces to
\begin{equation} 
J_0\bigg|_{\epsilon = 0} \;=\;
{16 \over 3 T^4} \int_0^{\infty}dp \; {p^4 \over E_p} n(E_p)\;. 
\label{J0}
\end{equation}

\subsection{Two loops}\label{1.2}

The free energy at second order in $\bar g$  
comes from the two-loop diagram labeled 1a in Fig.~\ref{fig-F}, 
and also from inserting the order-$\bar g^2$ mass counterterm 
$\Delta_1 m^2$ into the one-loop diagram:
\bqa
{\cal F}_1 &=& {\cal F}_{\rm 1a}  
+ {\partial {\cal F}_0 \over \partial \bar m^2} \Delta_1 m^2\,.
\label{F1-def}
\eqa
The expression for the diagram 1a is
\bqa
{\cal F}_{\rm 1a} &=& {1 \over 8} \bar g^2 {\cal I}_1^2 \,,
\label{F1a-int}
\eqa
where the sum-integral ${\cal I}_n$ is defined in (\ref{In}).
Keeping only the temperature-dependent terms, 
this diagram is 
\bqa
{\cal F}_{\rm 1a} &=& 
{\bar \alpha \over 8 (4\pi)^2} \left( {\mu\over\bar m} \right)^{4\epsilon}
\left[ - 2 \left( {1\over\epsilon} + 1 + O(\epsilon) \right) J_1 \bar m^2 T^2
	+  J_1^2 T^4 \right] \,,
\label{F1a}
\eqa
where $\bar \alpha = \bar g^2/(4\pi)^2$.
The pole proportional to $J_1 {\bar m}^2 T^2$ is canceled by the last term 
in (\ref{F1-def}).  
The identity (\ref{jrecursion}) is useful for calculating 
the derivative ${\partial J_0/\partial {\bar m}^2}$ in that term.
The mass counterterm is thereby determined to be
\bqa
\Delta_1 m^2 &=& {1 \over 2 \epsilon} \, \bar \alpha \, \bar m^2\,.
\label{delm2-1}
\eqa
%
Our final result for the two-loop contribution to the free energy is
\bqa
{\cal F}_1 &=& {\bar \alpha \over 8 (4 \pi)^2}
\left[  - 2(\bar L+1) J_1 \bar m^2 T^2
	+ J_1^2 T^4 \right] \;,
\label{F1}
\eqa
where $\bar L=\log(\mu^2/ \bar m^2)$ 
and $J_1$ is the function of $\beta \bar m$
defined in (\ref{Jn}). In the limit $\epsilon \to 0$, it reduces to
\begin{equation} 
J_1\bigg|_{\epsilon = 0} \;=\;
{8 \over T^2} \int_0^{\infty}dp \; {p^2 \over E_p} n(E_p)\;. 
\label{J1}
\end{equation}

\subsection{Three loops}\label{3}

The free energy at second order in $g^2$ 
comes from the three-loop diagrams labeled 2a and 2b in Fig.~\ref{fig-F}, 
and also from inserting counterterms 
into the one-loop and two-loop diagrams:
\bqa
{\cal F}_2 &=& 
{\cal F}_{\rm 2a} + {\cal F}_{\rm 2b} 
+ {\,\partial {\cal F}_{\rm 1a} \over \partial \bar m^2}  \Delta_1 m^2 
+ {{\cal F}_{\rm 1a} \over \bar g^2} \Delta_1 g^2
+{1\over2} {\partial^2{\cal F}_0 \over (\partial \bar m^2)^2} 
	\left(\Delta_1 m^2\right)^2 
+ {\partial {\cal F}_0 \over \partial \bar m^2} \Delta_2 m^2 \,.
\label{F2-def}
\eqa
The expressions for the diagrams 2a and 2b are
\bqa
{\cal F}_{\rm 2a} &=& -{1\over16} \bar g^4 {\cal I}_1^2 {\cal I}_2  \; , 
\\
{\cal F}_{\rm 2b} &=&-{1\over48} \bar g^4 {\cal I}_{\rm ball} \; .
\eqa
Using the expressions for ${\cal I}_n$ in the Appendix 
and for ${\cal I}_{\rm ball}$ in (\ref{Iball-final}), the
temperature-dependent terms in these diagrams are
\bqa
{\cal F}_{\rm 2a} &=& 
{\bar \alpha^2 \over 16(4\pi)^2} \left({\mu \over \bar m}\right)^{6\epsilon}
\left[ 
- \left( {1\over\epsilon^2} + {2\over\epsilon} + 3 + {\pi^2\over6} \right)
	J_2 \bar m^4
+ 2 \left( {1\over\epsilon^2} + {1\over\epsilon} + 1 + {\pi^2\over6} \right)
	J_1 \bar m^2 T^2 \right.
\nonumber
\\
&& \hspace{3cm} 
\left. + 2\left( {1\over\epsilon} + 1 \right) J_1 J_2 \bar m^2 T^2
	- {1\over\epsilon} J_1^2 T^4
	- J_1^2 J_2 T^4 
\right] \,,
\label{F2a}
\\ 
{\cal F}_{\rm 2b} &=& 
{\bar \alpha^2 \over 16(4\pi)^2} \left({\mu \over \bar m}\right)^{6\epsilon}
\left[ 
\left(  {2\over\epsilon^2} + {17\over 3\epsilon} - {4 \over 3} C_1 \right) 
	J_1 \bar m^2 T^2 \right.
\nonumber
\\
&& \hspace{3cm} 
\left. - \left( {2\over\epsilon}+ 4 \right) J_1^2 T^4
	- \left( 2 K_2 + {4\over3} K_3 \right) T^4 
\right] \,.
\label{F2b}
\eqa

The identity (\ref{jrecursion}) is useful for computing the derivatives
with respect to ${\bar m}^2$ in (\ref{F2-def}).
The pole in $\epsilon$ proportional to $J_1^2 T^4$ in (\ref{F2a})
is canceled by the $\Delta_1 g^2$ term in (\ref{F2-def}).  
After taking into account the terms 
in (\ref{F2-def}) involving $\Delta_1 m^2$, the remaining poles 
in (\ref{F2a}) and (\ref{F2b}) are proportional to $J_1 \bar m^2 T^2$ 
and are canceled by the $\Delta_2 m^2$ term in (\ref{F2-def}). 
The new counterterms that enter at this order are
\bqa
\Delta_1 g^2 &=& {3 \over 2\epsilon} \, \bar \alpha  \, \bar g^2 \,,
\label{delg2-1}
\\
\Delta_2 m^2 &=& 
\left( {1\over2\epsilon^2} - {5\over24\epsilon} \right)
	\, \bar \alpha^2 \, \bar m^2 \,.
\label{delm2-2}
\eqa
Our final result for the three-loop contribution to the free energy is
\bqa
&& {\cal F}_2 \;=\;
{\bar \alpha^2 \over 16 (4\pi)^2}
\left[  - (\bar L+1)^2 J_2 \bar m^4
+ \left( 4 \bar L^2 + {28\over3} \bar L - 4 - {\pi^2\over3} 
	- {4\over3} C_1 \right) J_1 \bar m^2 T^2 
\right.
\nonumber
\\
&&\hspace{3cm}
\left.
	+ 2 (\bar L + 1) J_1 J_2 \bar m^2 T^2
	- (3 \bar L + 4) J_1^2 T^4
	- J_1^2 J_2 T^4 
	- \left( 2 K_2 + {4 \over 3} K_3 \right) T^4 
\right] ,
\label{F2}
\eqa
where $C_1 = -9.8424$.  The functions $K_2$, $K_3$, $J_0$, and $J_1$, 
are given by (\ref{K2}), (\ref{K3}), (\ref{J0}), and (\ref{J1}),  and $J_2$ is
\begin{equation} 
J_2\bigg|_{\epsilon = 0} \;=\;
4 \int_0^{\infty}dp \; {1 \over E_p} n(E_p)\;. 
\label{J2}
\end{equation}
The complete result for the free energy to order $\bar \alpha^2$
is the sum of (\ref{F0}), (\ref{F1}), and (\ref{F2}).

\section{Physical parameters}

In this section, we express the free energy in terms of the 
physical mass $m$ of the boson at zero temperature and 
the physical coupling constant $g$ defined by the threshold 
scattering amplitude at zero temperature.

\subsection{Physical mass}

The physical mass $m$ of the scalar particle at zero temperature 
is given by the location of the pole in the propagator.
If $\Pi(P^2)$ is the self-energy function in Euclidean space, 
then $m$ satisfies
\beq
P^2 + \bar m^2 + \Pi(P^2) = 0 \qquad {\rm at} \; P^2 = - m^2. 
\label{m2-eq}
\eeq
This equation can be solved perturbatively for $m^2$ as a function of
the parameters $\bar m$ and $\bar g$ defined by dimensional 
regularization and minimal subtraction.  To express the free energy 
in terms of $m$ to three-loop order, we need to calculate $m^2$ 
to order $\bar g^4$.

\begin{figure}
\epsfxsize=7cm
\centerline{\epsffile{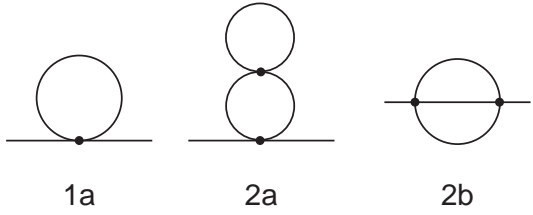 }}
\caption{One-loop and two-loop diagrams 
	that contribute to the self-energy.} 
\label{fig-Pi}
\end{figure}

The one-loop self-energy $\Pi_1$, which is independent of $P^2$,
can be written
\beq
\Pi_1 \;=\; \Pi_{\rm 1a} + \Delta_1 m^2 \, . 
\eeq
The expression for the one-loop diagram 1a in Fig.~\ref{fig-Pi} is
\beq
\Pi_{\rm 1a} \;=\; {1 \over 2} \bar g^2 I_1 \, , 
\eeq
where the one-loop integral $I_1$ is given in (\ref{i1}).
Adding the counterterm in (\ref{delm2-1}),
the one-loop self energy is 
\beq
\Pi_1 \;=\;  
	-{1\over2}(\bar L + 1) 
	\bar \alpha \, \bar m^2 \, .
\label{Pi1}
\eeq

The two-loop self-energy function, which depends on $P^2$,  is
\bqa
\Pi_2(P^2) &=&
\Pi_{\rm 2a} + \Pi_{\rm 2b}(P^2)
+ {\partial \Pi_{\rm 1a} \over \partial \bar m^2} \Delta_1 m^2
+ {\Pi_{\rm 1a} \over \bar g^2} \Delta_1 g^2 + \Delta_2 m^2 \,. 
\label{Pi2-def}
\eqa
The expressions for the two-loop diagrams 2a and 2b in 
Fig.~\ref{fig-Pi} are
\bqa
\Pi_{\rm 2a} &=&
- {1 \over 4} \bar g^4 I_1 I_2 \,,
\\
\Pi_{\rm 2b}(P^2) &=& - {1 \over 6} \bar g^4 I_{\rm sun}(P^2) \,. 
\eqa
To calculate the physical mass to order $\bar \alpha^2$,
we need the value of $\Pi_{\rm 2a}(P^2)$ only at $P^2 = - \bar m^2$.
Inserting the values for $I_1$ and $I_2$ from the Appendix
and the value of $I_{\rm sun}(-\bar m^2)$ from (\ref{Isun-final}),
we obtain
\bqa
\Pi_{\rm 2a} &=&
{\bar \alpha^2 \over 4}  \left({\mu \over \bar m}\right)^{4\epsilon}
\left[ {1 \over \epsilon^2} + {1 \over \epsilon} + 1 + {\pi^2 \over 6}
	\right] \bar m^2\,,
\\
\Pi_{\rm 2b}(- \bar m^2) &=& 
{\bar \alpha^2 \over 4}  \left({\mu \over \bar m}\right)^{4\epsilon}
\left[ {1 \over \epsilon^2} + {17 \over 6\epsilon} - {2 \over 3} C_1 \right] 
	\bar m^2 \,. 
\eqa
%
%
%
Combining all of the terms in (\ref{Pi2-def}), the value of the two-loop 
self-energy at $P^2 = -\bar m^2$ is 
\bqa
\Pi_2(- \bar m^2) &=& 
\left[ {1 \over 2} \bar L^2 + {7\over6} \bar L - {1\over2} - {\pi^2\over24} 
	- {1\over 6} C_1 \right] 
	\bar \alpha ^2  \, \bar m^2 \,. 
\label{Pi2}
\eqa

The solution to the equation (\ref{m2-eq}) for $m^2$ to order $\bar \alpha^2$ is
\beq
m^2 \;=\; \bar m^2  + \Pi_1 + \Pi_2(- \bar m^2) \, . 
\label{m2-sol}
\eeq
Inserting (\ref{Pi1}) and (\ref{Pi2}) into (\ref{m2-sol}), 
our final result for the physical mass to order $\bar \alpha^2$ is
\beq
m^2 \;=\; 
\left[ 1 -  {1 \over 2}(\bar L + 1) \bar \alpha 
	+ \left( {1 \over 2}\bar L^2 + {7 \over 6} \bar L - {1\over2} 
		- {\pi^2\over24} - {1\over6} C_1 \right)
	\bar \alpha^2 \right] \bar m^2 . 
\label{m2}
\eeq

\subsection{Physical coupling constant}

A convenient physical definition of the coupling constant $g$ is
that the amplitude for $2 \to 2$ scattering is exactly $- g^2$ 
at threshold where all four particles have four-momentum $p = (m, 0)$.
To express the free energy 
in terms of $g$ to three-loop order, we need to calculate $g^2$ 
to order $\bar g^4$. 

The one-loop expression for the negative of the
scattering amplitude at threshold is 
\beq
g^2 \;=\; \bar g^2 - {1 \over 2} \bar g^4 
\left[ I_{\rm bubble}(-4 m^2) + 2 \, I_{\rm bubble}(0) \right]
+ \Delta_1 g^2. 
\label{g2-int}
\eeq
where the bubble integral is defined in (\ref{Ibubble}).
Using the result (\ref{feynman}), the values of the bubble integrals
that appear in (\ref{g2-int}) are 
\bqa
I_{\rm bubble}(-4 m^2) &=& 
{1 \over (4 \pi)^2} \left( {\mu \over \bar m} \right)^{2 \epsilon} 
\left[ {1 \over \epsilon} + 2 \right] \,,
\label{Ibubble-1}
\\
I_{\rm bubble}(0) &=& 
{1 \over (4 \pi)^2} \left( {\mu \over \bar m} \right)^{2 \epsilon} 
{1 \over \epsilon}  \,.
\label{Ibubble-2}
\eqa
We have neglected the difference between $m$ and $\bar m$
in the bubble integrals because it is higher order in $\bar \alpha$.
Inserting (\ref{Ibubble-1}) and (\ref{Ibubble-2}) 
together with (\ref{delg2-1}) into (\ref{g2-int}),
our final result for the physical coupling constant 
$\alpha = g^2/(4 \pi)^2$ is
\beq
\alpha \;=\; 
\left[ 1 - \left({3\over2} \bar L + 1 \right) \bar \alpha \right] \bar \alpha \, . 
\label{g2}
\eeq

\subsection{Three-loop free energy}

To express the three-loop free energy in terms of the physical mass 
and coupling constant, we need to invert (\ref{m2}) and (\ref{g2})
to obtain $\bar m^2$ and $\bar g^2$ in terms of $m^2$ and $g^2$,
insert them into our expression for the free energy, 
and expand to order $\alpha^2$.
Inverting (\ref{m2}) and (\ref{g2}), we obtain
\bqa
\bar m^2 &=& 
\left[ 1 + {1\over2}(L + 1) \alpha 
	+ \left( {1 \over 2}L^2 + {1 \over 3} L + 1 + {\pi^2\over24} 
		+ {1\over6} C_1 \right) \alpha^2 
	\right] m^2 \,,
\label{mbar2}
\\
\bar \alpha &=& 
\left[ 1 + \left({3\over2} L + 1 \right) \alpha \right] \alpha \,, 
\label{gbar2}
\eqa
where $L = \log(\mu^2/m^2)$, not to be confused with 
$\bar L = \log(\mu^2/\bar m^2)$ in (\ref{m2}) and (\ref{g2}).
Upon inserting these expressions into the sum of (\ref{F0}), (\ref{F1}), 
and (\ref{F2}), and expanding to order $\alpha^2$, 
all the terms that depend on $L$ cancel.  Our final expression for the
temperature-dependent contribution to the free energy 
in terms of physical parameters is
\bqa
{\cal F} &=&
- {1 \over 2(4\pi)^2} 
\left[ J_0  
\;-\; {\alpha \over 4} J_1^2 
\;+\; {\alpha^2 \over 8}
	\left( 2 J_1^2 + J_1^2 J_2 + 2 K_2 + {4 \over 3} K_3 \right) 
\right] T^4 \,,
\label{finalresult}
\eqa
where $K_n$ and $J_n$ are the functions of $\beta m$
defined by (\ref{K2}), (\ref{K3}), (\ref{J0}), (\ref{J1}), and (\ref{J2}).
This expression is remarkably compact.
\begin{figure}
\epsfxsize=14cm
\centerline{\epsffile{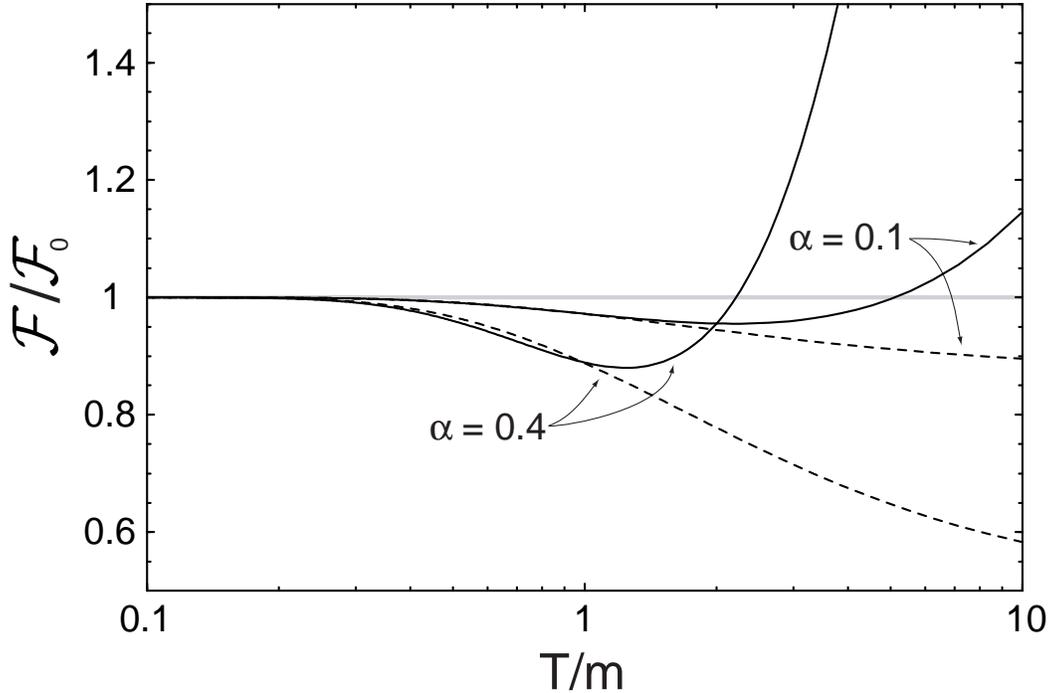}}
\vspace{5mm}
\caption{Free energy normalized to that of an ideal gas vs. 
	$T/m$ for $\alpha = 0.1$ and 0.4.
	The dashed and solid lines are the two-
	and three-loop approximations to the free energy, respectively.} 
\label{fig-PvsT}
\end{figure}

The effect of the interaction on the free energy
(which is the negative of the pressure) is illustrated in 
Fig.~\ref{fig-PvsT}.
We normalize the free energy to that of an ideal gas of particles 
with the same physical mass $m$, 
which is given by ${\cal F}_0$ in (\ref{F0}).
We plot ${\cal F}/{\cal F}_0$ as a function of $T/m$ on a log scale
for two different values of the physical coupling constant:  
$\alpha=0.1$ and $\alpha=0.4$, which correspond to
$g=3.97$ and $g=7.95$, respectively.
The dashed lines are the free energies truncated 
after the order-$\alpha$ terms.  The solid lines are the free energies 
truncated after the order-$\alpha^2$ terms.

For $T \ll m$, the three-loop result for the free energy 
(\ref{finalresult}) approaches 
\bqa
{\cal F} &\longrightarrow&
{\cal F}_0 
\left\{ 1 \;-\; 
{\alpha - 2 \alpha^2 \over 2} 
\left( {2 \pi T \over m} \right)^{1/2}
e^{-m/T}
\right\} \, .
\eqa
The exponential approach to the free energy of an ideal gas is
evident in Fig. \ref{fig-PvsT}.
Note that the order-$\alpha^2$ correction is smaller than the
order-$\alpha$ correction only if $\alpha < \mbox{${1 \over 2}$}$.
For $T \gg m$, (\ref{finalresult}) approaches
\bqa
{\cal F} &\longrightarrow&
{\cal F}_0 
\left\{ 1 \;-\; {5 \over 4} \alpha
\;+\; 
\left[ {5 \pi \over 4} {T \over m} - {15 \over 4} \log {T \over m} 
	- 6.6245 \right] \alpha^2
\right\} \,.
\eqa
In the order-$\alpha^2$ correction, the linearly divergent $T/m$ term
is the first of a series of infrared divergent terms
that behave like
$\alpha^{n+1} (T/m)^{2n-1}$.  
These terms come from the ring diagrams which, when summed
to all orders, give a correction of $+(5 \sqrt{6}/3) \alpha^{3/2}$. 
The logarithm in the order-$\alpha^2$ correction term
arises from the running of the coupling constant.
It can be absorbed into the order-$\alpha$ correction term by replacing
the physical coupling constant $\alpha$ by $\bar \alpha(T)$,
the $\overline{MS}$ coupling constant with renormalization scale $\mu=T$. 
For $T\gg m$, we expect the three-loop result to be a good approximation only if the
$\alpha^2 T/m$ correction is small compared to the $\alpha$ correction,
which requires $T\ll m/(\pi \alpha)$.

\section{Summary}

We have reduced the thermal basketball diagram for a massive scalar 
field theory with a $\phi^4$ interaction to three-dimensional integrals 
that can be evaluated numerically.  As an application, we
calculated the free energy for this theory to order $\alpha^2$.  
The result is particularly simple if the free energy is expressed 
in terms of the physical mass and coupling constant.
Another useful application of our result for
the massive thermal basketball diagram 
would be to extend the calculation of the free energy for the massless theory 
using screened perturbation theory to three-loop accuracy \cite{A-B-S}.

\section*{Acknowledgments}
This work was supported in part by the U.~S. Department of 
Energy Division of High Energy Physics 
(grants DE-FG02-91-ER40690 and DE-FG03-97-ER41014)
and by a Faculty Development Grant 
from the Physics Department of the Ohio State University.
Two of us (J.O.A.~and E.B.) would like to thank the Institute
for Nuclear Theory at the University of Washington for their
hospitality during the initial phase of this project.

\appendix\bigskip\renewcommand{\theequation}{\thesection.\arabic{equation}}

\section{One-loop Sum-integrals}
\setcounter{equation}{0}

The one-loop sum-integrals required to calculate the free energy to
order $g^4$ are
\bqa
{\cal I}_0' &=& - \sumint_P \log \left( P^2 + m^2 \right) \,,
\label{I0'}
\\
{\cal I}_n &=& \sumint_P {1 \over (P^2+m^2)^n} \,.
\label{In}
\eqa
The sum-integral ${\cal I}_0'$ is the derivative  of ${\cal I}_n$
with respect to its index evaluated at $n=0$.
These integrals satisfy
\bqa
{\partial \ \ \over \partial m^2}{\cal I}_0' &=& - {\cal I}_1 \,,
\\
{\partial \ \ \over \partial m^2} {\cal I}_n &=& - n {\cal I}_{n+1} \,.
\eqa

The specific sum-integrals that are required are ${\cal I}_0'$,
${\cal I}_1$, and ${\cal I}_2$.  The temperature-dependent terms
in the sum-integrals can be conveniently expressed 
in terms of the following integrals: 
\begin{equation} 
J_n \;=\;
{ 4 e^{\gamma\epsilon} \Gamma({1\over2}) 
	\over \Gamma({5\over2}-n-\epsilon) }
{m^{2 \epsilon} \over T^{4-2n}}
\int_0^{\infty}dp \; {p^{4-2n-2\epsilon} \over E_p} n(E_p)\;. 
\label{Jn}
\end{equation}
These integrals are functions of $\beta m$ only
and satisfy the  recursion relation
\bqa
m {\partial \ \over \partial m} J_n \;=\;
2 \epsilon J_n - 2 (\beta m)^2 J_{n+1}\;.
\label{jrecursion}
\eqa
If we separate out the temperature-dependent terms in the one-loop 
sum-integrals, the resulting expressions are
\begin{eqnarray}
{\cal I}_0' &=&
{1\over(4\pi)^2} \left({\mu\over m}\right)^{2\epsilon}
\left[ {e^{\gamma\epsilon} \Gamma(1+\epsilon) 
	\over \epsilon (1-\epsilon) (2-\epsilon)} m^4
+ J_0 T^4 \right]\;,
\label{si0}
\\
{\cal I}_1 &=&
{1\over(4\pi)^2} \left({\mu\over m}\right)^{2\epsilon}
\left[ - {e^{\gamma\epsilon} \Gamma(1+\epsilon) 
	\over \epsilon (1-\epsilon)} m^2
+ J_1 T^2 \right]\;,
\label{si1}
\\
{\cal I}_2 &=&
{1\over(4\pi)^2} \left({\mu\over m}\right)^{2\epsilon}
\left[ {e^{\gamma\epsilon} \Gamma(1+\epsilon) \over \epsilon}
+ J_2 \right]\;.
\label{si2}
\eqa

To calculate the physical mass and coupling constant,
we also need the one-loop Euclidean momentum integrals
$I_1$ and $I_2$ defined by
\beq
I_n \;=\; \int_P {1 \over (P^2+m^2)^n}\;.
\eeq
These integrals satisfy
\beq
{\partial \ \ \over \partial m^2} I_n \;=\; - n I_{n+1}\;.
\eeq
The integrals $I_1$ and $I_2$ are identical to the 
temperature-independent terms in (\ref{si1}) and (\ref{si2}), respectively.
Expanding around $\epsilon=0$, these integrals are  
\begin{eqnarray}
I_1 &=&
{1\over(4\pi)^2} \left({\mu\over m}\right)^{2\epsilon}
\left[ - {1 \over \epsilon} - 1 
	- \left( 1 + {\pi^2\over 12} \right) \epsilon + O(\epsilon^2)  \right] 
	m^2 \,,
\label{i1}
\\
I_2 &=&
{1\over(4\pi)^2} \left({\mu\over m}\right)^{2\epsilon}
\left[ {1\over\epsilon} + {\pi^2\over12} \epsilon + O(\epsilon^2) \right] \,.
\label{i2}
\eqa

\vspace{1.5cm}
\centerline{\bf Addendum}
\vspace{.75cm}

After this paper was completed, J.-M. Chung \cite{Chung-Private} provided us
with analytic expressions for the coefficients $C_0$ and $C_1$ defined
in (\ref{I0-final}) and (\ref{Isun-final}):
\begin{eqnarray}
C_0 &=& {275 \over 12} + {23 \over 2}\,\zeta(2) - 2\,\zeta(3) \nonumber \, , \\
C_1 &=& -{59 \over 8} -  {3 \over 2}\,\zeta(2) \, . \nonumber
\end{eqnarray}
These analytic expressions can be derived using the methods described in Refs.
\cite{Groote,Chung-ChungANDBerends}.

\end{document}